\documentclass[aps,prl,reprint,amsmath,amssymb,superscriptaddress]{revtex4-2}

\usepackage{graphicx}
\usepackage{bm}
\usepackage{hyperref}
\usepackage{color}
\usepackage{easyReview}
\usepackage{latexsym} 
\usepackage{ulem}

\begin{document}

\title{Anomalous acousto-current within the quantum Hall plateaus}

\author{Renfei Wang}
\affiliation{International Center for Quantum Materials,
  Peking University, Haidian, Beijing, China, 100871}
  \author{Xiao Liu}
\affiliation{International Center for Quantum Materials,
  Peking University, Haidian, Beijing, China, 100871}
\author{Mengmeng Wu}
\affiliation{International Center for Quantum Materials,
  Peking University, Haidian, Beijing, China, 100871}

\author{Yoon Jang Chung} 
\affiliation{Department of Electrical Engineering, 
	Princeton University, Princeton, New Jersey, USA, 08544} 
\author{Adbhut Gupta} 
\affiliation{Department of Electrical Engineering, 
	Princeton University, Princeton, New Jersey, USA, 08544}
\author{Kirk W. Baldwin} 
\affiliation{Department of Electrical Engineering, 
	Princeton University, Princeton, New Jersey, USA, 08544}
\author{Mansour Shayegan} 
\affiliation{Department of Electrical Engineering, 
	Princeton University, Princeton, New Jersey, USA, 08544}
\author{Loren Pfeiffer} 
\affiliation{Department of Electrical Engineering, 
	Princeton University, Princeton, New Jersey, USA, 08544}

\author{Xi Lin} 
\email{xilin@pku.edu.cn}
\affiliation{International Center for Quantum Materials, 
  Peking University, Haidian, Beijing, China, 100871}
\affiliation{Hefei National Laboratory, Hefei, China, 230088}

\author{Yang Liu} 
\email{liuyang02@pku.edu.cn}
\affiliation{International Center for Quantum Materials, 
  Peking University, Haidian, Beijing, China, 100871}
\affiliation{Hefei National Laboratory, Hefei, China, 230088}

\date{\today}

\begin{abstract}

  We systematically study the acousto-current of two-dimensional electron systems in the integer and fractional quantum Hall regimes using surface acoustic waves. We are able to separate the co-existing acoustic scattering and drag, when phonons induce drag current and tune the electron conductivity, respectively. At large acoustic power, the drag current is finite when the system is compressible and exhibits minima when incompressible quantum Hall effects appear. Surprisingly, it exhibits anomalously large bipolar spikes within the quantum Hall plateaus while it vanishes linearly with reduced acoustic power at compressible phases. The current peaks reverse their polarity at the two flanks of exact integer or fractional fillings, consistent with the opposite electric charge of the quasiparticle/quasihole.
    
\end{abstract}
\maketitle

Ultra-high mobility two-dimensional electron systems (2DES) exhibit quantum Hall
effect\cite{IQH} when subjected in high perpendicular magnetic field $B$. 
An incompressible superfluid phase that can host
dissipationless current at extremely low temperature $T$
\cite{Tsui.PRL.1982, theQHE, Jain.CF.2007}. Their quasiparticle
excitations are anyons with topological phases and have attracted
tremendous interests, especially those of the 5/2 fractional quantum
Hall state who might obey non-Abelian statistics and be useful for
topological quantum computing
\cite{Willett.PRL.1987,Nayak.RevModPhys.2008}.  Varies experimental
techniques are employed to study the quasi-particles's properties,
including weak tunneling\cite{Roddaro.PRL.2003, Miller.Nature.2007,
  Radu.Science.2008, Fu.PNAS.2016},
interferometry\cite{Willett.PNAS.2009, Zhang.PRB.2009,
  Willett.PRB.2010, Nakamura.Nature.2020}, shot
noise\cite{Saminadayar.PRL.1997, Picciotto.PHYSICAB.1998,
  Bid.Nature.2010, Dolev.PRB.2010} and thermal
transport\cite{Chickering.PRB.2013, Banerjee.Nature.2017,
  Banerjee.Nature.2018}, etc.

\begin{figure}[htbp]
	\includegraphics{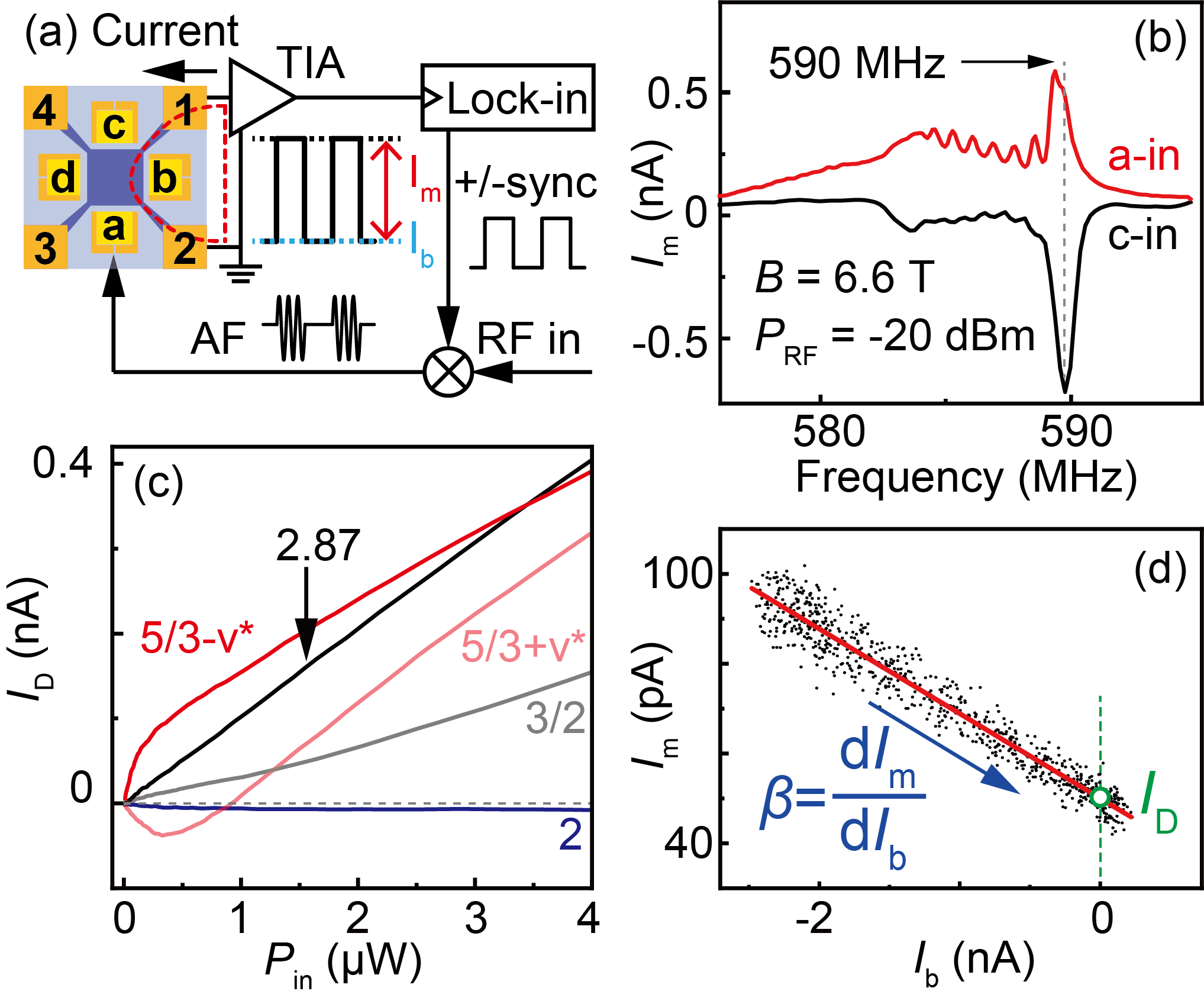}
	\caption{\label{fig1} (a) A schematic drawing of our experiment. The
		2DES mesa (dark blue) has van der Pauw geometry with four contacts
		(labeled 1-4). We evaporated four IDTs (labeled a-d) for launching
		acoustic waves. An on-off modulated (at about 10 Hz) excitation
		voltage is applied to the emitting IDT. The current captured by the
		TIA at zero excitation power is the bias current
		($I_{\text{b}}$). The current flowing through the sample changes by
		acousto-current $I_{\text{m}}$ when the acoustic wave is
		launched. We use lock-in technique to measure $I_\text{m}$ and a
		multimeter to measure $I_\text{b}$. (b) The measured $I_\text{m}$
		vs.  excitation frequency using IDT a and c. (c) The measured drag
		current $I_\text{D}$ at $I_\text{b}=0$ as a function of input
		excitation power $P_{\text{in}}$ at different $\nu$. (d) Typical
		$I_\text{m}$ vs.  $I_\text{b}$. The SAW generates a drag current
		$I_\text{D}$ and affects the 2DES conductance modeled by scattering
		$\beta=\text{d}I_\text{m}/\text{d}I_\text{b}$.}
\end{figure}

Surface acoustic wave (SAW) is a useful technique which studies 2DES
through the electron-phonon interaction \cite{Wixforth.PRL.1986,
  Wixforth.PRB.1989, Willett.PRL.1990, Paalanen.PRB.1992,
  Willett.PRL.1993, Shilton.PRB.1995, Simon.PRB.1996.SAWTheory}. It is
a special acoustic wave that propagates along the interface between
materials with different acoustic velocity. In piezoelectric materials
such as GaAs, SAW can be launched by applying an excitation voltage to
the emitter interdigital transducer (IDT) and detected by measuring
the induced voltage on the receiver IDT \cite{White.1965.IDT}. In
these materials, the electron-phonon interactions are enhanced by
the piezoelectric field accompanying the mechanical
vibration. Studying quantum phenomena in 2DES through the acoustic
propagation properties, i.e. its attenuation and velocity shift, is a
heavily exploited technique \cite{Willett.PRL.2002, Friess.PRL.2018,
  Friess.PRL.2020, Friess.Nature.2017, Drichko.PRB.2011,
  Drichko.PRB.2016, Fang.PRL.2023.GrapheneSAWOscillation}. The
electron-phonon interaction can also transfer momentum from
propagating phonon to electrons and induce a current. Pioneering
studies have observed the flow of such phonon drag currents (or
equivalently the accumulation of longitudinal voltages) under various
experimental conditions \cite{Esslinger.1992.Drag,
  Esslinger.1994.Drag, Shilton.1995.AEEffect, Rampton.1996.AE.2DHS,
  Kennedy.1998.AE.2DES2DHS, Rotter.1998.GiantAEEfect,
  Dunford2002.doublelayer1, Dunford2002.doublelayer2,
  Miseikis.2012.Graphene, Bandhu.2013.Graphene, ZhaoPai.2020.APL,
  ZhaoPai.2022.PRL}.


\begin{figure*}[htbp]
	\includegraphics{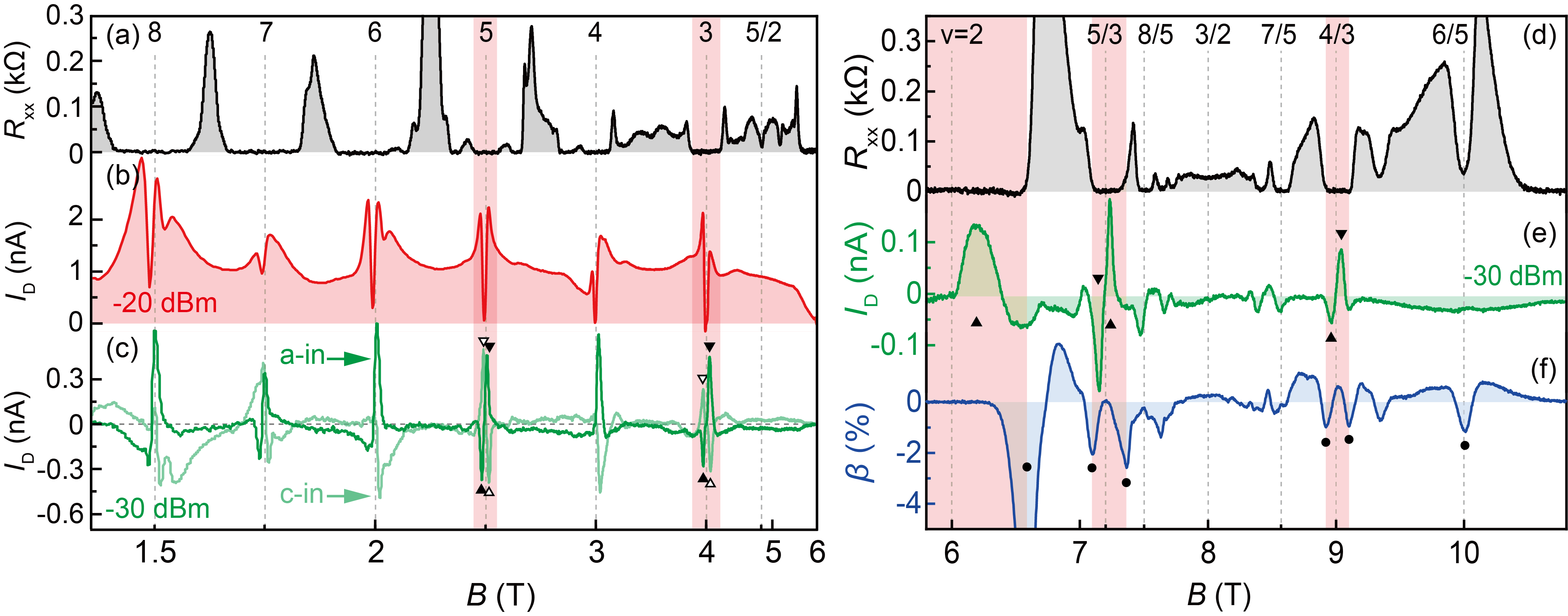}
	\caption{\label{fig2} (a \& d) Longitudinal magnetoresistance
		$R_{\text{xx}}$ taken by standard transport measurement in the
		absence of SAW. The current flows between contacts 3 \& 4, and the
		voltage is measured between contacts 1 \& 2. (b) The drag current
		$I_\text{D}$ taken at $P_{\text{in}}=$ - 20 dBm using IDT a. (c \&
		e) $I_\text{D}$ taken at $P_{\text{in}}= - 30$ dBm. The thick green
		curve is taken using IDT a, and the light green curve uses IDT
		c. (f) The relative conductivity change $\beta$ at $P_{\text{in}}=$
		- 40 dBm with IDT a.}
\end{figure*}

In this study, we conducted a systematic and thorough investigation of
the acousto-current in both the integer and fractional quantum Hall
regime. After carefully optimizing the measurement, we are able to
separate two co-existing phenomena: the phonon-drag effect that
describes the current driven by acoustic waves and the
phonon-scattering effect when the 2DES conductivity is tuned by the acoustic
waves. When the input acoustic power is orders of magnitude lower than
previous studies, we discover an anomalously large phonon drag current
within the conduction plateaus at quantum Hall effects
surprisingly. This phenomenon exists even at the fragile 5/2 quantum
Hall effects.

In this work, we investigate an ultra-high mobility 2DES confined in
30-nm-wide GaAs quantum well whose density is
$n= 2.91 \times 10^{11} \mathrm{cm^{-2}} $ and low temperature
mobility is $ \mu = 2 \times 10^7 \mathrm{cm^2/(V \cdot s)} $. The
2DES mesa (dark blue region in Fig. \ref{fig1}(a)) has a Van der Pauw
geometry with four evaporated Au/Ge/Ni/Au contacts (gold square
labeled 1 to 4). Two pairs of 5-$\mathrm{\mu}$m-period IDTs evaporated
on each side (gold square labeled a to d) converts the input voltage
into SAW propagating along the $\left \langle 110\right \rangle $ crystal direction\footnote{ All
  doping layers underneath the IDTs are removed by wet etching to
  avoid unwanted signal coupling. The typical attenuation of our
  delay-line device is approximately - 30 dB.}. We parameterize the
acoustic power by the input microwave power $P_{\text{in}}$. The
typical attenuation of our delay-line device is approximately - 30
dB\cite{wu2023morphing}. The current is measured by a compact, battery powered
trans-impedance amplifier (TIA) with carefully designed current loop
to achieve $ \lesssim 10 \text{fA} $ leak current \cite{SI}. A lock-in
scheme is used to measure the acousto-current $I_{\text{m}}$ from the
TIA output voltage; see Fig. 1(a) and its caption. The measurements
are carried out in a dilution refrigerator whose base temperature is
below 20 mK.

The SAW consists a continuous flow of monochromatic phonons and
interacts with 2DES by exchanging momentum and energy via
electron-phonon interaction. At $P_{\text{in}}=$ - 20 dBm, the process
of transferring momentum from phonons to electrons, i.e. phonon drag,
dominates. Fig. \ref{fig1}b shows that the measured acousto-current
has a peak at about 590 MHz when the excitation frequency is close to
the IDT resonance frequencies $f=\lambda/v$ ($\lambda$ is the spatial
period of the IDT and $v$ is the SAW velocity). The dragged electrons
move along the same direction with the phonons (which is from IDT a to
c), so that a positive current flows into the sample from contact
1. The current changes its polarity if we reverse the phonon direction
by exciting IDT c instead of a; see Fig. \ref{fig1}(b). Furthermore,
the drag current is proportional to $P_{\text{in}}$, see
Fig. \ref{fig1}c. All these observations agree with previous reports
where up to $\mu$A drag currents is obtained using
$P_{\text{in}}\gtrsim - 10$ dBm \cite{Esslinger.1992.Drag,
  Bandhu.2013.Graphene, Miseikis.2012.Graphene}.

In Fig. \ref{fig2}(a), the longitudinal resistance $R_{\text{xx}}$
exhibits zero plateaus at integer quantum Hall effect, as well as
clear minima corresponding to fractional quantum Hall effects at
Landau level filling factor $\nu=5/2$, 7/3, etc. Fig. \ref{fig2}(b)
shows the drag current $I_\text{D}$ measured at $P_{\text{in}}=$ - 20
dBm. $I_\text{D}$ has minima at integer $\nu$, consistent with the
incompressibility of quantum Hall liquid. The $I_\text{D}$ minima at
integer $\nu$ in Fig. \ref{fig2}(b) are considerably narrower than the
$R_{\text{xx}}$ plateaus (highlighted by the red shading), while
previous studies suggest that $I_\mathrm{D}$ should be similar with
$R_{\text{xx}}$ \cite{Falko.PRB.47.9910, Esslinger.1994.Drag,
  Dunford2002.doublelayer1}. This is possibly related to the fact that
the electrons' mean free path in our ultra-high mobility samples
($\sim100$ $\mu$m) is much larger than the 5 $\mu$m acoustic
wavelength. A rigorous theoretical treatment is necessary.

The $P_{\mathrm{in}}=$ - 20 dBm used in Fig. \ref{fig2}(b) is
comparable with, if not much lower than, any previous studies. The
drag current $I_\mathrm{D}\sim 1$ nA is much smaller than typical
current used in transport measurements. Note that $I_{\mathrm{D}} = 1$
nA means that SAW carries $ I_{\mathrm{D}}/\text{e}f \approx 10 $
electrons per cycle. The $I_\mathrm{D}$ vs. $P_{\mathrm{in}}$ is roughly linear
in Fig. \ref{fig1}(c) at most fillings. The fact that $I_\mathrm{D}$
is always positive evidences that the particles being dragged by SAW
always have negative charge. The estimated induced charge density is
approximately $10^6$ cm$^{-2}$ (see the supplementary material of
Ref. \cite{wu2023morphing}), much smaller than the thermal fluctuation
k$T*\pi\hbar/m^*\sim 10^{7}$ cm$^{-2}$ at 10 mK. However, even such a
small excitation might already be too large for studying fragile
phases.


\begin{figure}[htbp]
	\includegraphics{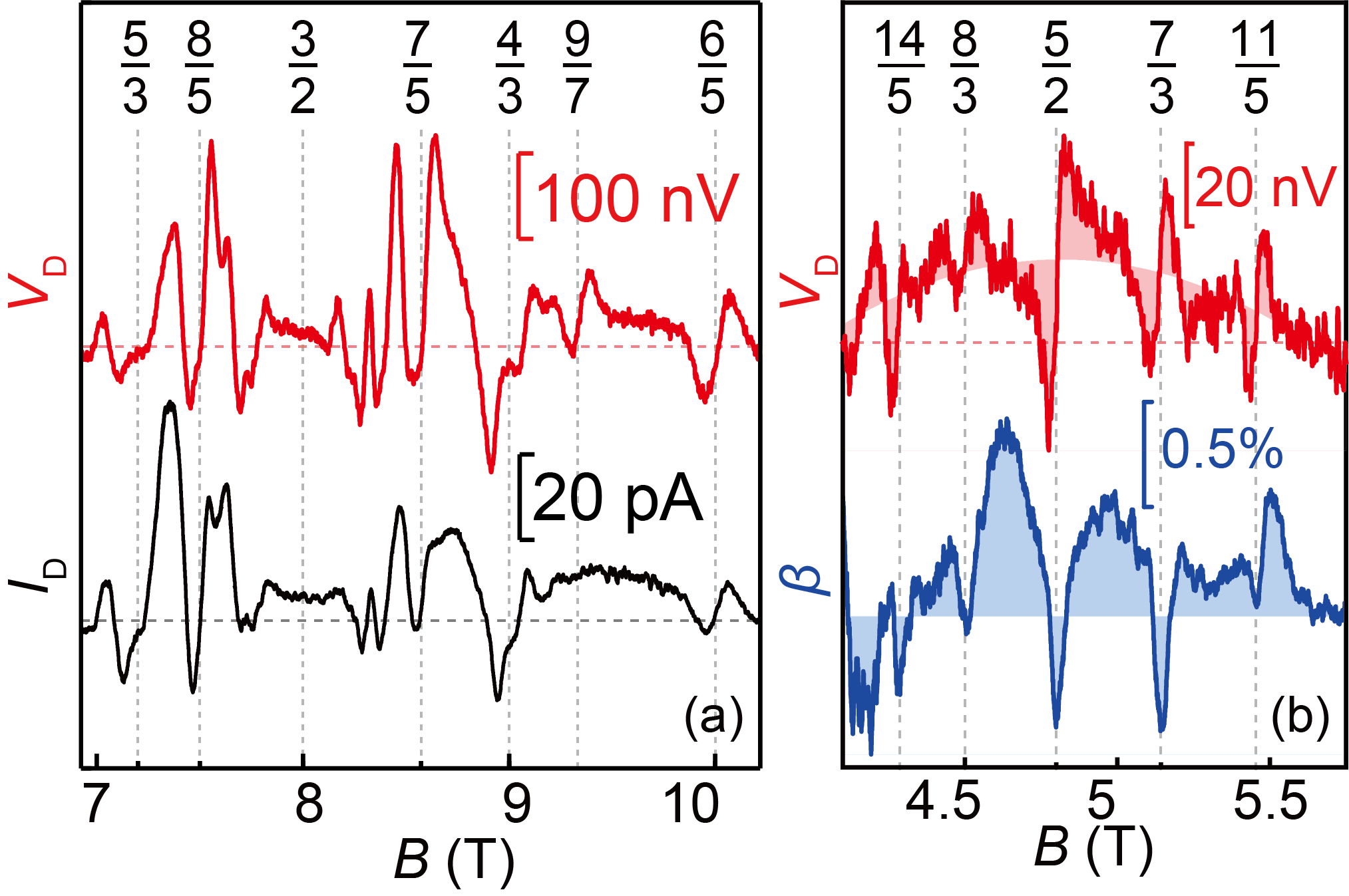}
	\caption{\label{fig3} (a) Drag voltage $V_\mathrm{D}$ and current $I_\mathrm{D}$
		measured near $\nu=3/2$. (b) $V_\mathrm{D}$ and $\beta$ near $\nu=5/2$. All
		data are measured using IDT a. $V_\mathrm{D}$ is taken at $P_{\mathrm{in}}= - 50$
		dBm. $I_\mathrm{D}$ \& $\beta$ are taken at $P_{\mathrm{in}}= - 40$ dBm.}
\end{figure}

The $I_\text{D}$ vs. $P_{\text{in}}$ relation near exact integer or fractional
filling becomes nonlinear when we reduce the excitation power to below
1 $\mu$W (-30 dBm); see Fig. \ref{fig1}(c). Note that $\lesssim 100$ pA
$I_\mathrm{D}$ corresponds to less than 1 electrons per each acoustic
period moving from contact 2 to contact 1! The measured
$I_\mathrm{D}$ in Fig. \ref{fig2}(c) becomes dramatically different
from the Fig. \ref{fig2}(b) data. $I_\mathrm{D}$ decreases to nearly
zero when the 2DES is compressible, agreeing with the linear $I_\text{D}$
vs. $P_{\text{in}}$ dependence in Fig. \ref{fig1}(c). Surprisingly, in
contrast to the $I_\mathrm{D}$ minima seen at integer $\nu$ in
Fig. \ref{fig2}(b), we observe large current peaks in
Fig. \ref{fig2}(c) within the $R_{\mathrm{xx}}$ plateaus (highlighted
by the red shading). The polarity of the current peak reverses when we
switch the SAW direction, confirming the phonon-drag-origin of these
peaks.

It is quite interesting that these current peaks have a bipolar
feature, namely $I_\mathrm{D}=0$ at exact integer $\nu$ and it has a
pair of positive and negative peaks at slightly non-integer
$\nu$. This feature suggests that they are related to the momentum
transfer from phonons to the quasiparticles/quasiholes, which appear
at slightly larger/smaller $\nu$ (lower/higher $B$) and have
negative/positive electric charge. This bipolar feature is clear at
odd fillings $\nu=3$, 5, etc., and the high-filling-side peaks at even
fillings $\nu=2$, 4, 6, etc. disappear \cite{SI}.  This is likely
related to the fact that the Fermi level at odd fillings resides
between Landau levels separated by the small spin splitting
$g^*\mu_BB$, while it jumps between Landau levels separated by
$\hbar\omega_C$ at even fillings. The large Fermi energy discontinuity
at even integer $\nu$ leads to stronger confining potential for
localized quasiparticles.

This bipolar current peaks also appear at fractional quantum Hall
effects. Fig. \ref{fig2}(e) shows $I_\mathrm{D}$ measured at
$P_{\mathrm{in}}=$ - 30 dBm and $1<\nu<2$. $I_\mathrm{D}$ is less than
10 pA when electrons form compressible composite Fermion Fermi sea at
$\nu=3/2$. Large and clear bipolar current peaks appear within the
$R_{\mathrm{xx}}$ plateaus at the $\nu=4/3$ and 5/3 fractional quantum
Hall effect. If we further reduce $P_{\mathrm{in}}$ to - 40 dBm, the
bipolar current peaks becomes visible at $\nu=8/5$ and 6/5, see
Fig. \ref{fig3}(a).

A schematical model was proposed to describe the origin of the drag
current by electron-phonon interaction
\cite{Esslinger.1994.Drag,Simon.PRL.2001}. The phonons interact with
2DES through the piezoelectric field $E_\text{eff}$ which is screened
by the induced spacial electron density fluctuation
$\Delta n(\textbf{r},t)$.  The conductivity $\sigma$ has a dependence
on the local particle density, i.e. $\text{d}\sigma/\text{d}n\ne 0$,
which leads to a correction to the local current density
$\text{d}\sigma/\text{d}n \cdot \Delta n(\textbf{r},t)
E_\text{eff}(\textbf{r},t)$.  The time averaging of this correction
term contributes a DC current proportional to
$\text{d}\sigma/\text{d}n$. It is quite suggestive to observe a
similarity between $I_\text{D}$ and $d\eta/dB$ in the fractional
quantum Hall regime at $B\gtrsim 6$ T, since the SAW velocity shift
$\eta=\delta v/v$ depends on $\sigma[n(\textbf{r},t)]$
\cite{SI}. However, the fact that large and sharp $I_\text{D}$ peaks
appear only within strong quantum Hall plateaus, suggests that the
presence of Wigner crystal and superfluid quantum Hall liquid is
important. A microscopic model describing the dynamics of these
quantum phases would be necessary to understand our observations.

The phonons not only induce a drag current by transferring momentum
to the electrons, it also changes the sheet conductivity by scattering
moving electrons. The latter mechanism induces a scattering current
when a DC current $I_\mathrm{b}$ flows through the sample. Both the
drag and scattering current are induced by phonons, and will be part
of the measured signal $I_\mathrm{m} = I_\mathrm{D} + \beta \cdot I_\mathrm{b}$.
The scattering current is
naturally proportional to the DC bias current $I_\mathrm{b}$ and is
independent of the phonon propagation direction \cite{SI}, while the
drag current is independent of $I_\mathrm{b}$ and aligns with the
phonon moving direction. The typical voltage offset imposed to the
contacts by the instrument's trans-impedance preamplifiers is
$\sim100$ $\mu$V (e.g. SR830), so that $I_\mathrm{b}$ is of the order
of 10 nA. The scattering current, which is usually a few percent of
$I_\mathrm{b}$, becomes non-negligible for detecting sub-nA
$I_\mathrm{D}$.

We are able to tune $I_\mathrm{b}$ of our TIA and measure its value
precisely down to pA-level. Fig. \ref{fig1}(d) shows typical
$I_\text{m}$ vs. $I_\text{b}$ taken at $B\simeq$ 7.35 T, where
$I_\text{m}$ has a linear dependence on $I_\mathrm{b}$ through
$I_\mathrm{m} = I_\mathrm{D} + \beta \cdot I_\mathrm{b}$. The
intercept at $I_\mathrm{b}=0$ is the drag current $I_\mathrm{D}$, and
its slope $\beta=\mathrm{d}I_\mathrm{m}/\mathrm{d}I_\mathrm{b}$ is the
SAW-induced relative conductivity change. We can apply a periodic
tuning of $I_\mathrm{b}$ and extract $\beta$ from the corresponding
$I_\mathrm{m}$ oscillation using lock-in technique. Alternatively, we
can also deduce $\beta$ and $I_\mathrm{D}$ from a proper linear
combination of $I_\mathrm{m}$ measured at different $I_\mathrm{b}$.
Results obtained by these two methods are equivalent\cite{SI}.


\begin{figure}[htbp]
	\includegraphics{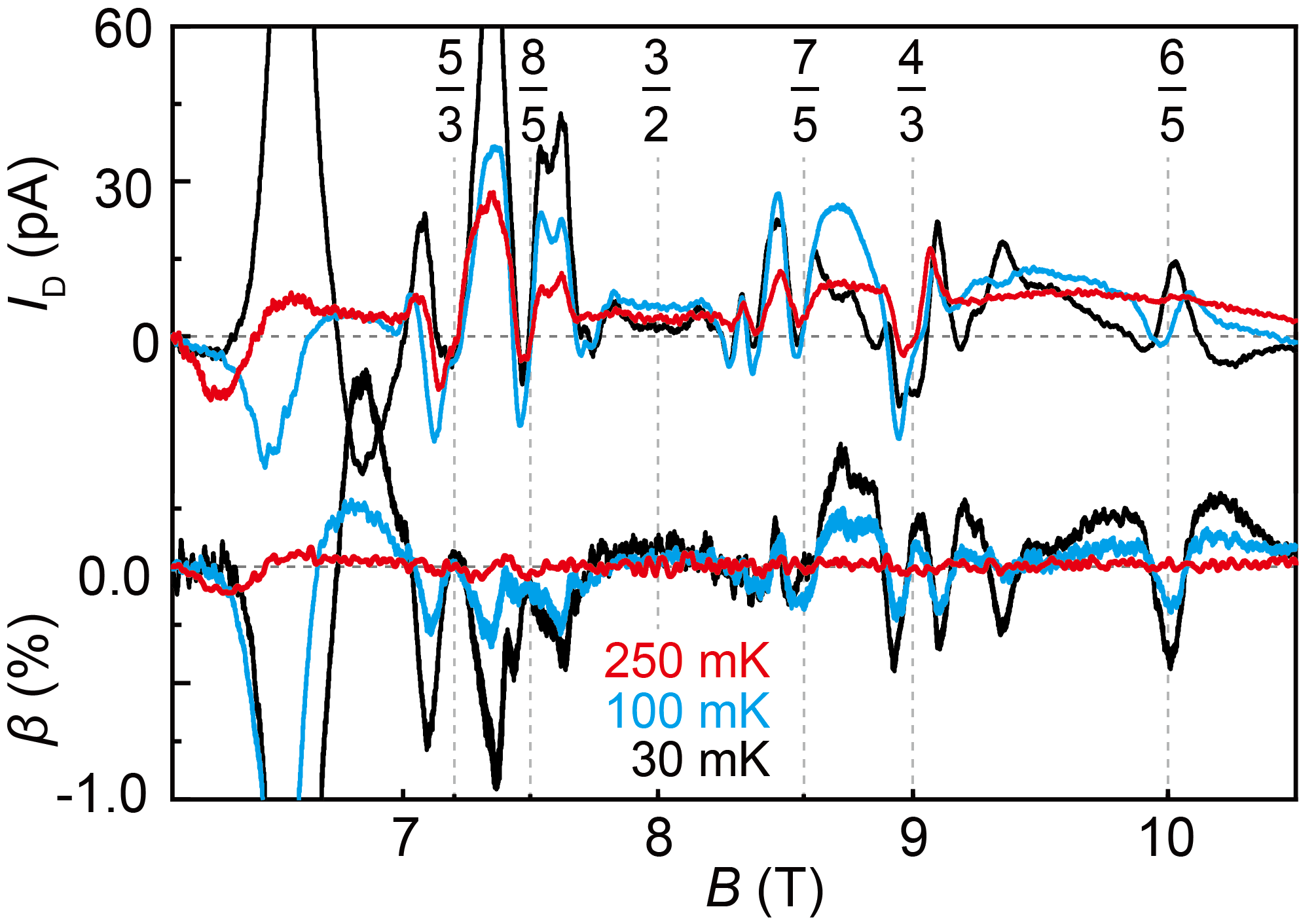}
	\caption{\label{fig4} $I_\mathrm{D}$ and $\beta$ near $\nu=3/2$ taken at
		different temperatures $T=$ 30, 100, and 250 mK. Data are measured
		using IDT a and $P_{\mathrm{in}}= - 40$ dBm.}
\end{figure}

Figure \ref{fig2}(f) shows $\beta$ measured at $1<\nu<2$ using
$P_{\mathrm{in}}= - 40$ dBm.  $\beta$ is also close to zero near
$\nu=3/2$. At strong fractional quantum Hall effects such as $\nu=$5/3
and 4/3 as well as the integer quantum Hall effects at $\nu=2$,
$\beta$ returns to zero at exact fillings, and exhibits two negative
peaks at the edge of the $R_{\mathrm{xx}}$ plateau, marked by black
dots in Fig. \ref{fig2}(f). At weak quantum Hall effects such as
$\nu$=6/5 and 9/7 where $R_{\mathrm{xx}}$ has no plateau, $\beta$
develops one single negative peak. In short, the $\beta$ peaks mark
the collapse of quantum Hall plateaus.

The bias current $I_\mathrm{b}$ is kept below 200 pA, so that the
scattering current becomes comparable with the $I_\mathrm{D}$ $\sim 10$ pA 
when $P_{\mathrm{in}}= - 40$ dBm.  We can eliminate the
scattering effect by measuring the open-loop voltage when there's no
current flowing through the sample ($I_\mathrm{b}\lesssim 1$ pA). In
this geometry, the charge accumulation at the upper and lower edge of
the mesa generates an electric field to compensate the momentum
transferred from the phonons. The measured open loop voltage
$V_\mathrm{D}$ between the two contacts is expected to be proportional
to the close loop current $I_\mathrm{D}$ through
$V_\mathrm{D}/I_\mathrm{D}=R_\mathrm{S}$, where $R_\mathrm{S}$ is the
resistance between the two contacts. Fig. \ref{fig3}(a) shows
$V_\mathrm{D}$ measured between contacts 1 and 2 at
$P_{\mathrm{in}}= - 50$ dBm . It is nearly identical to the
$I_\mathrm{D}$ measured with $P_{\mathrm{in}}= - 40$ dBm. Such an
excitation level is sufficiently low, so that we are able to observe
clear bipolar features in $V_\mathrm{D}$ and minima in $\beta$ that
corresponds to fragile quantum Hall effects at $\nu=5/2$ and 7/3, 8/3,
11/5 and 14/5; see Fig. \ref{fig3}(b).

A theoretical investigation of the drag and scattering effect is still
missing, therefore, it is important to describe the experimentally
observed features \cite{SI}. Firstly, $I_\mathrm{D}$ is an odd
function of quasiparticle filling factor $\nu^*$ \footnote{The
  quasiparticle/quasihole filling is $\nu^*=\nu-N$ for integer quantum
  Hall effects at integer filling $N$, and $\nu^*=\nu^{\mathrm{CF}}-p$
  for fractional quantum Hall effect at composite Fermion filling
  factor $p$; $N$ and $p$ are integers. The relation between electron
  filling factor and composite Fermion filling factor is
  $\nu=2-\nu^{\mathrm{CF}}/ \left( 2 \nu^{\mathrm{CF}}\pm 1 \right) $.} and has
opposite sign at the two flanks of quantum Hall effects, while $\beta$
is an even function of $\nu^*$ and its two peaks flanking the quantum
Hall effect are both negative. Secondly, the $I_\mathrm{D}$ peaks
appear within the $R_{\mathrm{xx}}$ plateaus at $ \nu^* \simeq 0.05 $,
and the $\beta$ peaks appears at the edge of the plateaus where
$\nu^* \sim 0.2$. Thirdly, the polarity of $I_\mathrm{D}$ depends on
the direction of SAW, while $\beta$ remains the same for all SAW
propagation direction. Finally, $I_\mathrm{D}$ and $\beta$ have
different dependence on temperature $T$. Fig. \ref{fig4} shows
$I_\mathrm{D}$ and $\beta$ taken with $P_{\mathrm{in}}= - 40$ dBm at
different $T$. As $T$ increases, $\beta$ weakens and completely
vanishes to zero at 250 mK. $I_\mathrm{D}$ remains finite up to 250 mK
although its bipolar features gradually disappear.

In conclusion, we investigate the SAW-induced current in an ultra-high
mobility 2DES. We find that the current response is a superposition of
both drag and scattering current. We observe anomalously large bipolar
peaks in drag current/voltage at quantum Hall effects, a strong
evidence for the interaction between phonon and
quasiparticles/quasiholes. By reducing the SAW power to orders of
magnitudes lower than previous studies, we are able to investigate
phases ranging from strong integer quantum Hall effects to the fragile
fractional quantum Hall effects at $\nu=5/2$.

\begin{acknowledgments}
  
 We acknowledge support by the National Key Research and Development Program of China (Grant No. 2021YFA1401900) and the National Natural Science Foundation of China (Grant No. 12074010 $ \& $ 12141001) for sample fabrication and measurement. The Princeton University portion of this research is funded in part by the Gordon and Betty Moore Foundation’s EPiQS Initiative, Grant GBMF9615.01 to Mansour Shayegan and Loren Pfeiffer. We thank Steve Simon for valuable discussions.
  
\end{acknowledgments}

\bibliography{20240226_SAWdrag_v1.4.bib}

\clearpage
\renewcommand*{\thefigure}{S\arabic{figure}}
\setcounter{figure}{0}

%
%
%

\title{Anomalous acousto-current within the quantum Hall plateaus}

\author{Renfei Wang}
\affiliation{International Center for Quantum Materials,
	Peking University, Haidian, Beijing, China, 100871}
\author{Xiao Liu}
\affiliation{International Center for Quantum Materials,
	Peking University, Haidian, Beijing, China, 100871}
\author{Mengmeng Wu}
\affiliation{International Center for Quantum Materials,
	Peking University, Haidian, Beijing, China, 100871}

\author{Yoon Jang Chung} 
\affiliation{Department of Electrical Engineering, 
	Princeton University, Princeton, New Jersey, USA, 08544} 
\author{Adbhut Gupta} 
\affiliation{Department of Electrical Engineering, 
	Princeton University, Princeton, New Jersey, USA, 08544}
\author{Kirk W. Baldwin} 
\affiliation{Department of Electrical Engineering, 
	Princeton University, Princeton, New Jersey, USA, 08544}
\author{Mansour Shayegan} 
\affiliation{Department of Electrical Engineering, 
	Princeton University, Princeton, New Jersey, USA, 08544}
\author{Loren Pfeiffer} 
\affiliation{Department of Electrical Engineering, 
	Princeton University, Princeton, New Jersey, USA, 08544}

\author{Xi Lin} 
\affiliation{International Center for Quantum Materials, 
	Peking University, Haidian, Beijing, China, 100871}
\affiliation{Hefei National Laboratory, Hefei, China, 230088}

\author{Yang Liu} 
\email{liuyang02@pku.edu.cn}
\affiliation{International Center for Quantum Materials, 
	Peking University, Haidian, Beijing, China, 100871}
\affiliation{Hefei National Laboratory, Hefei, China, 230088}

\date{\today}
	
\maketitle
\section{Supplementary Materials}
\subsection{Sample Information and measurement setup}

\begin{figure*}[htbp]
	\includegraphics{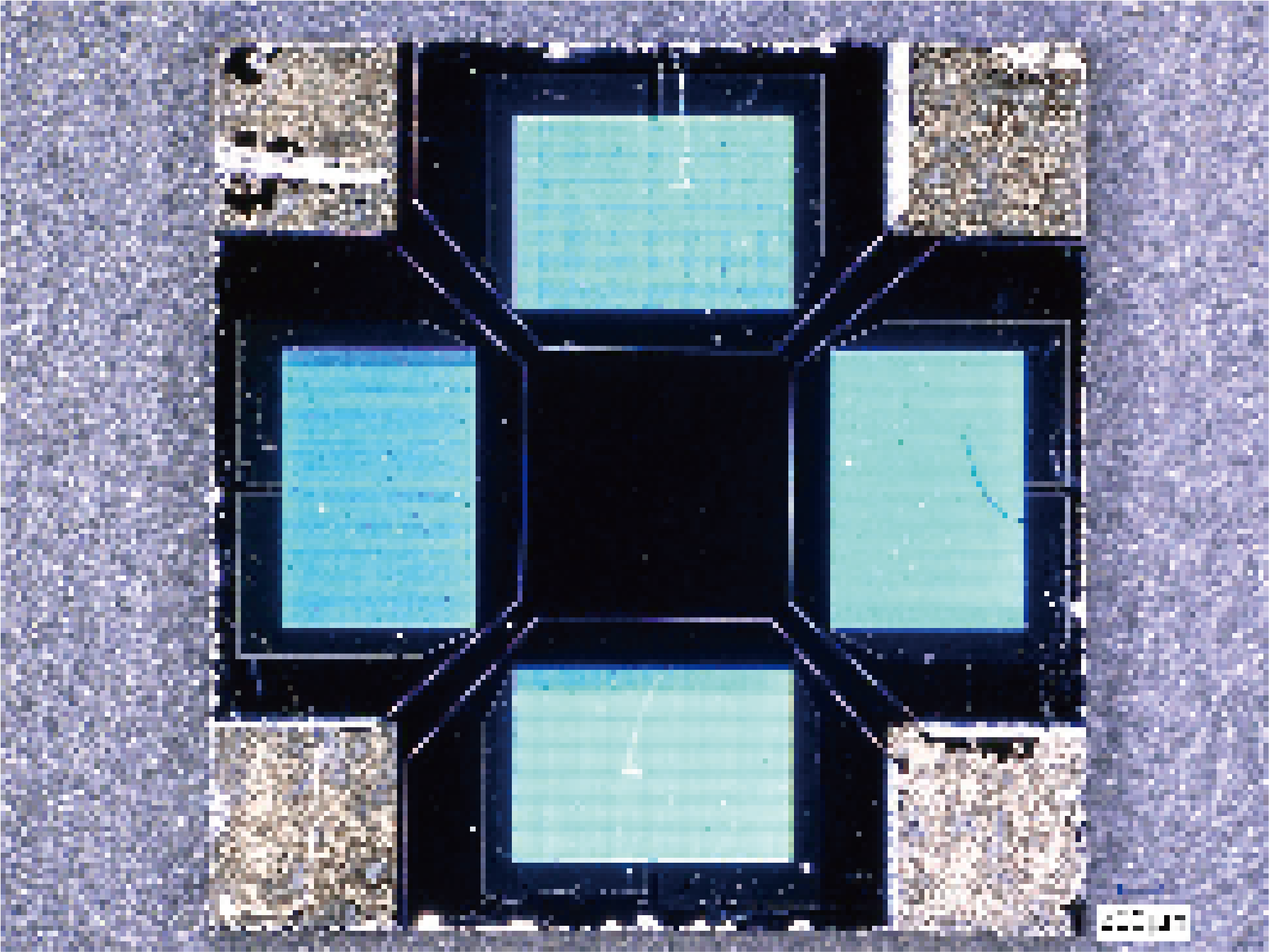}
	\caption{Photo of the sample. The entire sample has a side length of approximately 4 mm.}
	\label{SI_Fig_Sample}
\end{figure*}

Our sample is an ultra-high mobility 2DES confined in 30-nm-wide GaAs/AlGaAs quantum well grown by molecular beam epitaxy. The photo of sample is shown in Fig.\ref{SI_Fig_Sample}.
The Van der Pauw mesa is a square with a side length of 1.2 mm. And four 5 $\mathrm{\mu}$m-period interdigital transducers (IDTs) are evaporated outside of the mesa. All doping layers underneath the IDTs are removed by wet etching to avoid unwanted signal coupling. After all these sample fabrications, the typical attenuation of our delay-line device is approximately - 30 dB.

SAW propagates along the $ \langle 110 \rangle $ crystal direction. The large $R_{\text{xx}}$ peaks seen at $ \nu = $ 9/2, 11/2, etc. shown in Fig.2(b) of the manuscript suggest that the hard axis of these unidirectional charge density waves is along the SAW propagating direction.

\begin{figure*}[htbp]
	\includegraphics{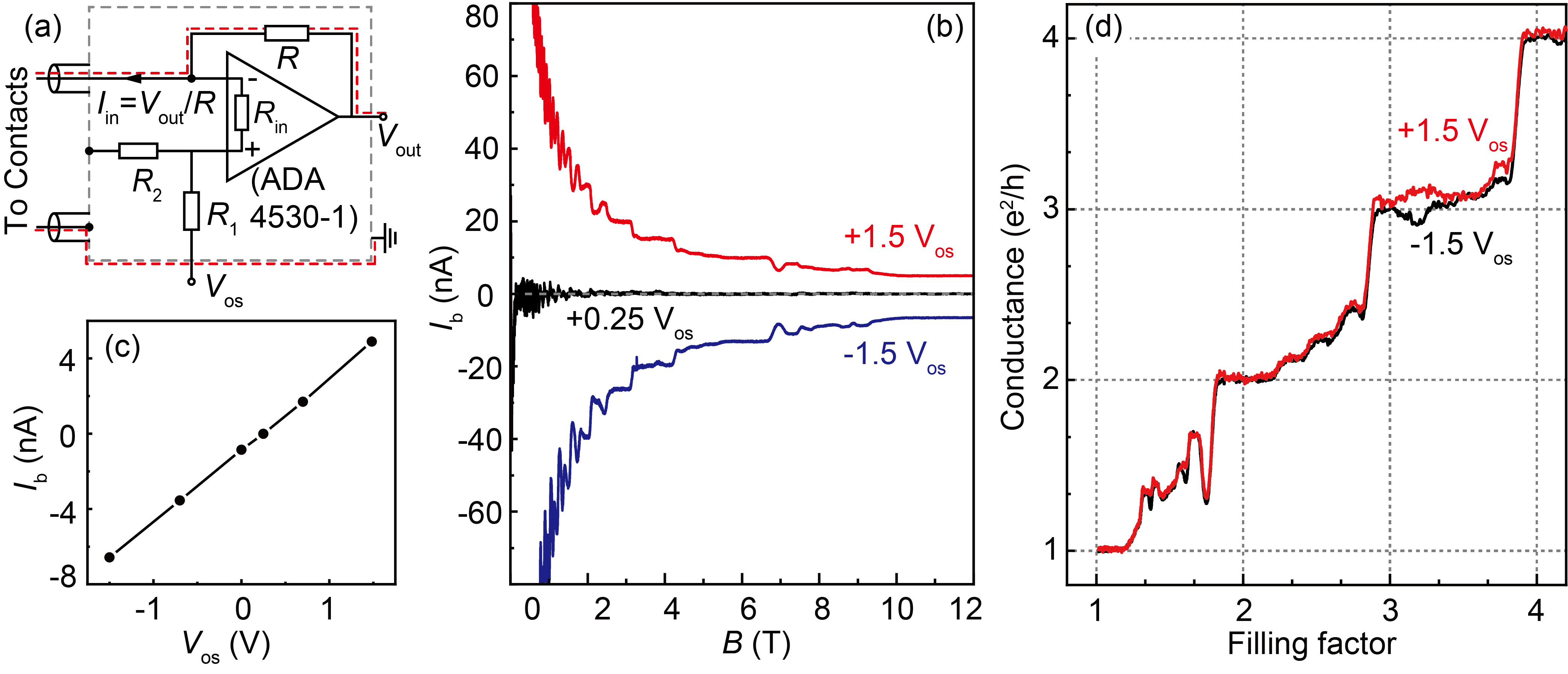}
	\caption{(a) Diagram of our compact trans-impedance amplifier. (b) The measured DC current traces for $V_{\mathrm{OS}}=+1.5$ V (red), +0.25 V (black) and -1.5 V (blue). (c) Linear relation between $V_{\mathrm{OS}}$ and $I_{\mathrm{b}}$ at $\nu=1$. (d) The two-point conductance of the sample calculated from $I_{\mathrm{b}}$ and $V_{\mathrm{OS}}$ for $V_{\mathrm{OS}}=+1.5$ V (red) and -1.5V (black).}
	\label{SI_Fig1}
\end{figure*}

We use operational amplifier (OPA) ADA4530 to build the compact trans-impedance amplifier (TIA) for current measurement, as shown in Fig.\ref{SI_Fig1}(a). This particular OPA has an extremely low leak current (about less than 10 fA). And the TIA input impedance $\lesssim 10$ $\mathrm{\Omega}$ in order to efficiently drain the drag current out of the sample. Combined with the usage of coaxial wires, it forces all current flowing out of the contact to pass through the feedback resistor $R$. So the TIA output voltage is precisely proportional to the current passing through the sample. As a result, the following relationship is precisely established:

\begin{eqnarray}
	V_{\mathrm{out}} = I_{\mathrm{in}} \times R
	\label{Vout}
\end{eqnarray}

When configured within its linear feedback regime, the potential difference between the inverting and noninverting inputs of an ideal OPA is zero. Unfortunately, this low-leaking current OPA have a rather large input offset voltage (up to $\pm$40 $\mathrm{\mu}$V). This implies that if we simply ground the noninverting input, a DC offset voltage will be applied at the input contact which induces a finite bias current $I_{\mathrm{b}}$ flowing into the sample. Therefore, we incorporate a voltage divider circuit ($R_{\mathrm{1}}$ = 100 k$\mathrm{\Omega}$, $R_{\mathrm{2}}$ = 10 $\mathrm{\Omega}$ ) at the noninverting input in order to adjust $I_{\mathrm{b}}$. Subsequently, we measure the TIA output voltage $V_{\mathrm{out}}$ to monitor the current through the sample. The measured DC current components for different Vos are displayed in Fig.\ref{SI_Fig1}(b). 

The magnitudes of the current at position $\nu$=1 with different $V_{\mathrm{OS}}$ is shown in Fig.\ref{SI_Fig1}(c). From these current and the applied $V_{\mathrm{OS}}$ we can deduce the two-point resistance of the sample using the following relation:
\begin{eqnarray}
	R_{\mathrm{2-point}} = (A \times V_{\mathrm{OS}} - B)/ I_{\mathrm{b}}
	\label{R2point}
\end{eqnarray}

where $A=R_{\mathrm{2}}/(R_{\mathrm{1}}+R_{\mathrm{2}})=10^{-4}$ is the voltage divider ratio and B is the input offset voltage of the OPA. From the linear relation between $V_{\mathrm{OS}}$ and $I_{\mathrm{b}}$ in Fig.\ref{SI_Fig1}(c), we can derive the sample’s 2-point resistance $R_{\mathrm{2-point}}=26.1$ k$\mathrm{\Omega}$ and the OPA’s input offset voltage (zero current) $B$ = 22.9 $\mu$V. It is consistent with the fact that the two-point resistance at $\nu=1$ is almost precisely $h/e^2=25.8$ k$\mathrm{\Omega}$ if the contact and wire resistance is small. 

The two-point conductance of the sample as a function of filling factor is obtained using Eq.\eqref{R2point} and shown in Fig.\ref{SI_Fig1}(d). The quantization of conductance occurs at integer and fractional fillings confirms the accuracy of our TIA.

In summary, our TIA enables precise monitoring of sample current with negligible leaking. We can introduce a bias voltage onto the sample contact and simultaneously measure the bias current, while employing lock-in technique to measure the AC current $I_{\mathrm{m}}$ induced by amplitude-modulated SAW.

\subsection{Two methods for obtaining the drag and scattering current}

SAW induced current $I_{\mathrm{m}}$ consists of two parts, the SAW drag $I_{\mathrm{D}}$ and the SAW scattering $\beta \cdot I_{\mathrm{b}}$, which can be separated using $I_{\mathrm{m}}$ measured at different bias current $I_{\mathrm{b}}$. We have mentioned that $I_{\mathrm{m}}$ has a linear dependence on $I_{\mathrm{b}}$ by the equation:
\begin{eqnarray}
	I_{\mathrm{m}} = I_{\mathrm{D}} + \beta \cdot I_{\mathrm{b}}
	\label{Im}
\end{eqnarray}
The SAW drag $I_{\mathrm{D}}$ is the $I_{\mathrm{m}}$ when there is no bias current flowing through the sample and $\beta=\text{d}I_\text{m}/\text{d}I_\text{b}$. We have two methods for deducing $I_{\mathrm{D}}$ and  $\beta$. 
Firstly, we can calculate $I_{\mathrm{D}}$ and $\beta$ using linear combination of $I_{\mathrm{m}}$ measured with different $I_{\mathrm{b}}$ through the flowing equations:
\begin{eqnarray}
\begin{cases}
	I_{\mathrm{m}}^1 =  I_{\mathrm{D}} + \beta \cdot I_{\mathrm{b}}^1\\
	I_{\mathrm{m}}^2 =  I_{\mathrm{D}} + \beta \cdot I_{\mathrm{b}}^2
\end{cases}
	\label{Im2}
\end{eqnarray}

\begin{eqnarray}
	\begin{gathered}
	I_{\mathrm{D}} = \frac{  I_{\mathrm{m}}^1 \cdot I_{\mathrm{b}}^2 - I_{\mathrm{m}}^2 \cdot I_{\mathrm{b}}^1  } {I_{\mathrm{b}}^2 -I_{\mathrm{b}}^1} \\
	\beta = \frac{I_{\mathrm{m}}^2-I_{\mathrm{m}}^1}{I_{\mathrm{b}}^2 -I_{\mathrm{b}}^1}
	\end{gathered}
\label{ID_beta}
\end{eqnarray}
This approach will be quoted as the DC- $V_{\mathrm{OS}}$  procedure.

\begin{figure*}[htbp]
	\includegraphics{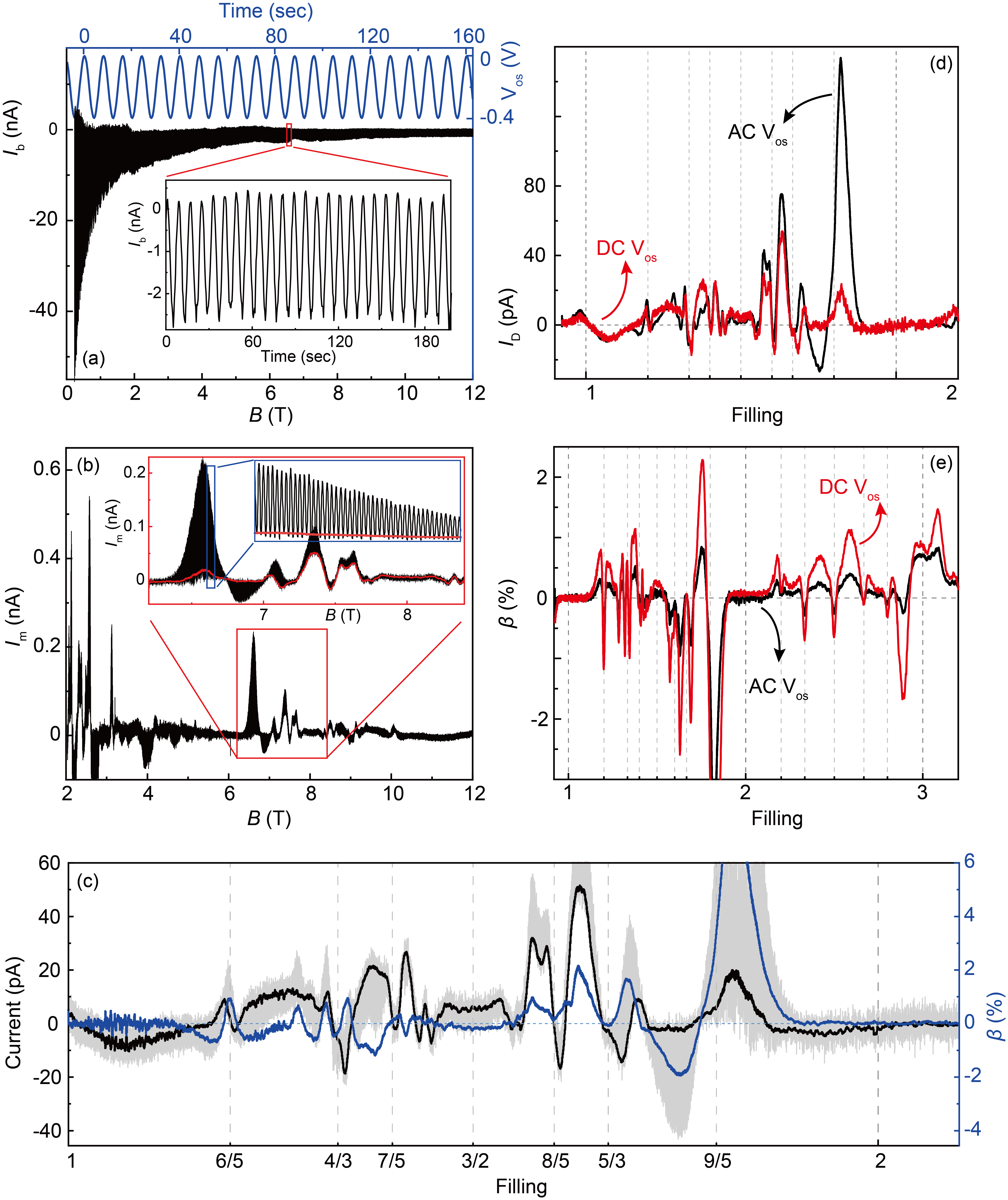}
	\caption{(a) The applied periodic voltage $V_{\mathrm{OS}}$ (blue curve at upper plate) and measured bias current $I_{\mathrm{b}}$ as a function of magnetic field for the AC-$V_{\mathrm{OS}}$ procedure. (b) The SAW induced current $I_{\mathrm{m}}$ consists an oscillating component and the red curve indicates the net drag current $I_{\mathrm{D}}$ at $I_{\mathrm{b}}= 0$. (c) The results of the raw $I_{\mathrm{m}}$ (gray background), $I_{\mathrm{D}}$ (black curve) and $\beta$ (blue with right axis). (d-e) The drag current $I_{\mathrm{D}}$ and $\beta$ obtained using two different methods are consistent with each other.}
	\label{SI_Fig2}
\end{figure*}

The second method, referred as AC-$V_{\mathrm{OS}}$ procedure, is applying an extremely low frequency (ten times smaller than the SAW modulation frequency) AC $V_{\mathrm{OS}}$:

\begin{eqnarray}
	V_{\mathrm{OS}} = V_{\mathrm{DC}} + V_{\mathrm{AC}} \cos(\omega_2 t)
	\label{ACVos}
\end{eqnarray}
Therefore, we have:
\begin{eqnarray}
	\begin{gathered}
		I_{\mathrm{b}} = I_{\mathrm{DC}} + I_{\mathrm{AC}}\cos(\omega_2 t)\\
		I_{\mathrm{m}} = I_{\mathrm{D}}(B) + \beta(B) [I_{\mathrm{DC}} + I_{\mathrm{AC}}\cos(\omega_2 t)]
	\end{gathered}
	\label{ACVosIm}
\end{eqnarray}
In our experiment, we vary $V_{\mathrm{OS}}$ between 0 and -0.4 V. The applied Vos and the measured $I_{\mathrm{b}}$ are shown in Fig.\ref{SI_Fig2}(a) and the $I_{\mathrm{m}}$ is shown in Fig.\ref{SI_Fig2}(b). We show the $I_{\mathrm{D}}= I_{\mathrm{m}} (I_{\mathrm{b}}=0)$ by the red traces in Fig.\ref{SI_Fig2}(b). 

Using lock-in technique, we can deduce $I_{\mathrm{AC}}$ from the TIA output, and $\beta(B) \cdot I_{\mathrm{AC}}$ from the oscillating component of measured $I_{\mathrm{m}}$, and then deduce $I_{\mathrm{D}}(B)$ by subtracting the scattering contribution $\beta(B) \cdot V_{\mathrm{OS}}$. The results of the raw $I_{\mathrm{m}}$, $I_{\mathrm{D}}$ and $\beta$ are shown in Fig.\ref{SI_Fig2}(c) using gray, black and blue curves, respectively. 
The $I_{\mathrm{D}}$ and $\beta$ obtained from the above two methods are consistent with each other, as shown in Fig.\ref{SI_Fig2}(d-e).

\subsection{Comparison of Bipolar drag current peaks with the SAW velocity shift}
\begin{figure*}[htbp]
	\includegraphics{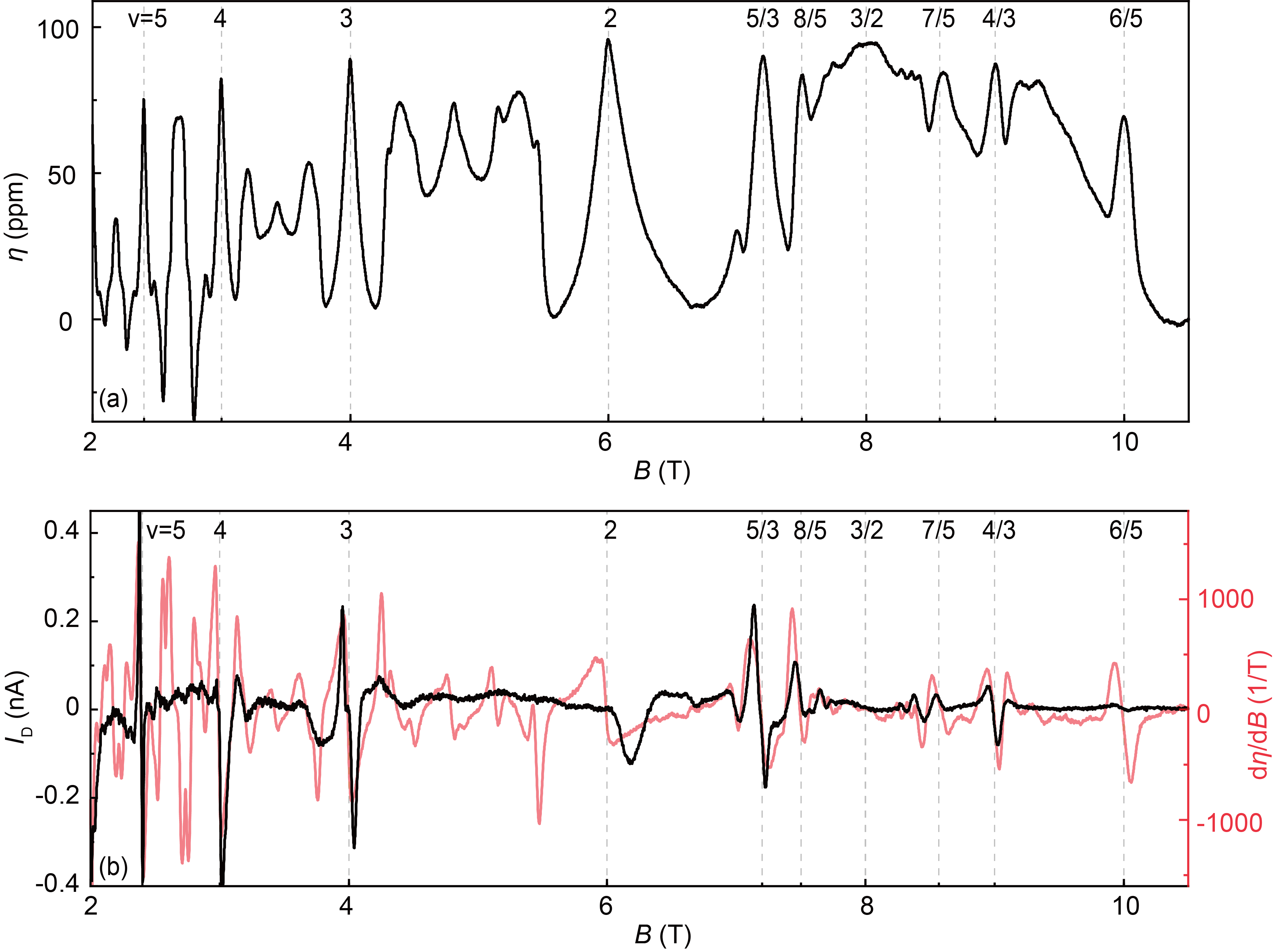}
	\caption{(a) The measured velocity shift $\eta = \Delta v/v$ from reference \cite{wu2023morphing}. (b) Comparison of the drag current $I_{\mathrm{D}}$ and the first derivative of $\eta$, $\mathrm{d}\eta/\mathrm{d}B$.}
	\label{SI_Fig5}
\end{figure*}

In a schematic model, the drag current can be express as:
\begin{eqnarray}
	J_\text{D}(\textbf{r},t) = [\sigma(n_\text{0}) + \text{d}\sigma/\text{d}n \cdot \Delta n(\textbf{r},t)] E_\text{eff}(\textbf{r},t)
	\label{JD}
\end{eqnarray}
In this model, the DC drag current induced by SAW is proportional to
$\text{d}\sigma/\text{d}n$. The measured velocity shift $\eta = \Delta
v/v$ cited from reference \cite{wu2023morphing} is shown in
Fig.\ref{SI_Fig5}(a). It is generally believed that the
bulk conductivity is the reason of this velocity shift. We take the
first derivative of $\eta$ ($\text{d}\eta/\text{d}B$) and compare it
with the drag current in Fig.\ref{SI_Fig5}(b). $\eta$ exhibits a V-shape peak and
discontinuity at exact fillings in its first derivative, which is
consistent with the presence of large bipolar values of
$\text{d}\sigma/\text{d}n$. The similarity between $I_{\mathrm{D}}$
and the $\text{d}\eta/\text{d}B$ is surprisingly good in fractional
states. However, the broad peaks of $\text{d}\eta/\text{d}B$ around
integer filling fill the entire plateaus, whereas the $I_{\mathrm{D}}$
peaks are narrower. Additionally, $\text{d}\eta/\text{d}B$ has rich
features at stripe phases and bubble phases, etc., while the
$I_{\mathrm{D}}$ only shows peaks within a narrow range around integer
quantum Hall plateaus. It is suggestive that the appearance of bipolar
$I_{\mathrm{D}}$ peaks may be related to the existence of Wigner
crystal. However, as of now, there is no detailed theoretical explanation available.

\subsection{The position of the drag current peaks}
\begin{figure*}[htbp]
	\includegraphics{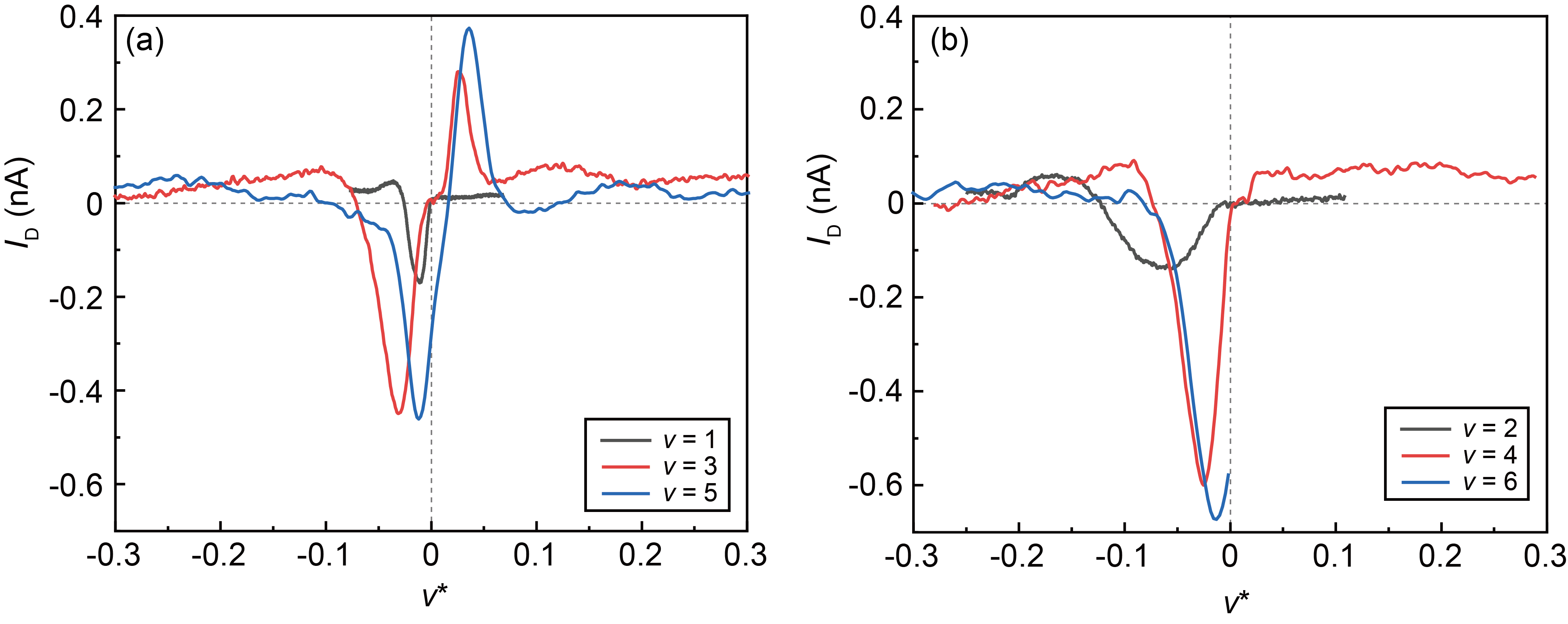}
	\caption{Drag current as a function of $\nu^*$ at the vicinity of (a) odd and (b) even integer filling.}
	\label{SI_Fig3}
\end{figure*}

The $I_{\mathrm{D}}$ peaks are observed within the Rxx plateaus at $\nu^* \approx 0.05 $. We define $\nu^* =\nu-\lfloor \nu \rfloor$ for integer filling. At odd filling, peaks at positive and negative $\nu^*$ are seen in Fig.\ref{SI_Fig3}(a). However, the positive-$\nu^*$ peaks disappear at even filling in Fig.\ref{SI_Fig3}(b). This might because that the single particle gap at positive $\nu^*$ of even filling is $\hbar \omega_c $ while it is $E_{\mathrm{z}}$ at odd fillings.

\subsection{The impact of SAW direction on current}
\begin{figure*}[htbp]
	\includegraphics{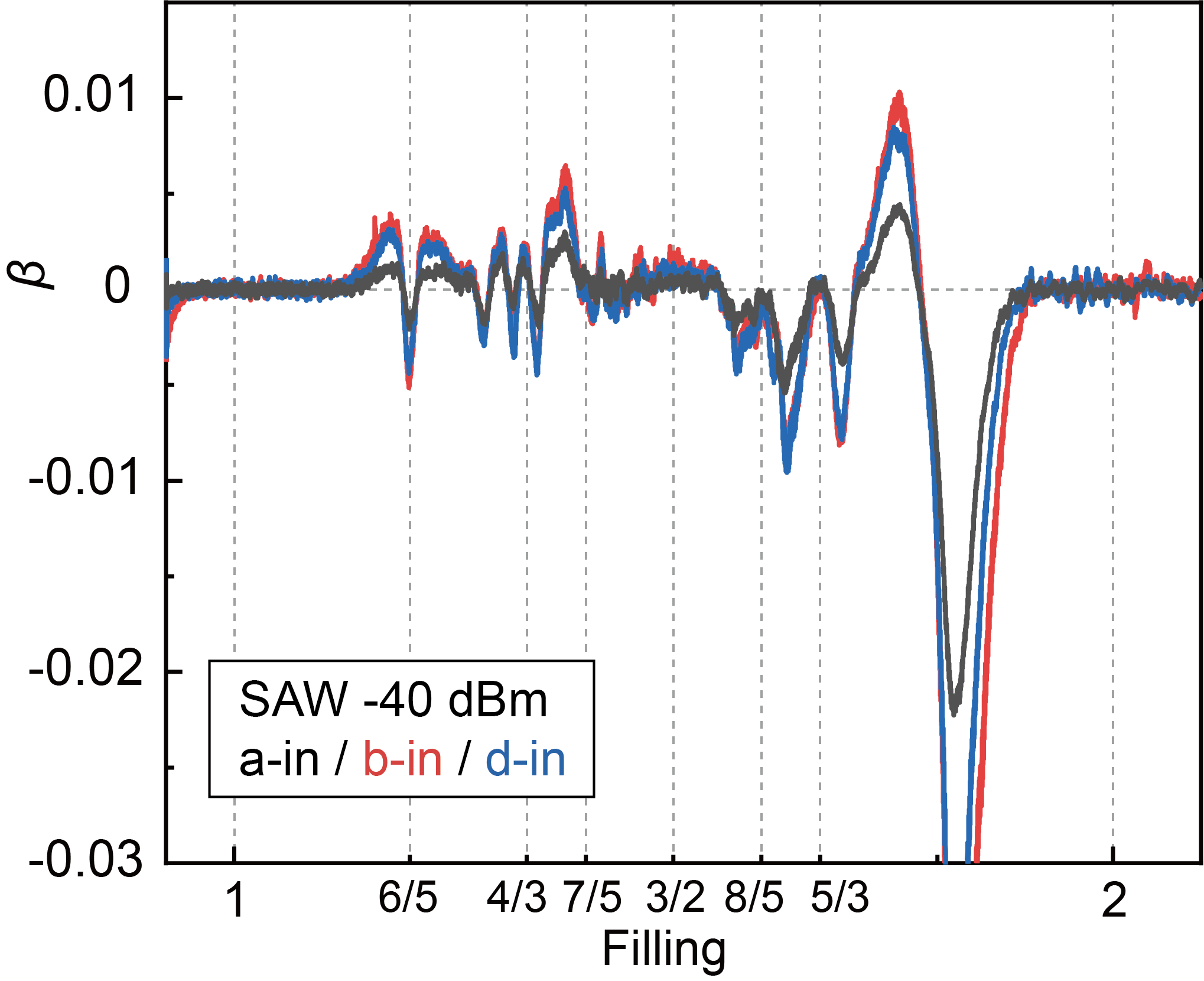}
	\caption{$\beta$ vs. filling when the SAW is launched from IDT-a (black), IDT-b (red) and IDT-d (blue). It is clear that $\beta$ is nearly independent with the SAW direction.}
	\label{SI_Fig4}
\end{figure*}

The SAW transforms the momentum to the carriers, resulting in a drag current. The polarity of this drag current flips when the SAW direction is reversed; see Fig.1(b) and 2(c) in the main text. The SAW also modulate the conductivity of the sample. The resulting scattering current is independent of the SAW propagation direction, see Fig.\ref{SI_Fig4}.

%

\end{document}